\documentclass[11pt,a4paper,english,superscriptaddress,aps]{revtex4}
\usepackage[latin9]{inputenc}
\usepackage[active]{srcltx}
\usepackage{babel}
\usepackage{amsmath}
\usepackage{amssymb}
\usepackage{graphicx}
\usepackage{esint}
\usepackage[normalem]{ulem}

\makeatletter

\@ifundefined{textcolor}{}
{%
 \definecolor{BLACK}{gray}{0}
 \definecolor{WHITE}{gray}{1}
 \definecolor{RED}{rgb}{1,0,0}
 \definecolor{GREEN}{rgb}{0,1,0}
 \definecolor{BLUE}{rgb}{0,0,1}
 \definecolor{CYAN}{cmyk}{1,0,0,0}
 \definecolor{MAGENTA}{cmyk}{0,1,0,0}
 \definecolor{YELLOW}{cmyk}{0,0,1,0}
}

\usepackage{graphicx}
\usepackage{babel}

\newcommand{\be}{\begin{equation}}
\newcommand{\ee}{\end{equation}}
\newcommand{\bea}{\begin{eqnarray}}
\newcommand{\eea}{\end{eqnarray}}

\makeatother

\begin{document}

\title{Chern-Simons modified gravity and closed time-like curves}

\author{P. J. Porf\'{i}rio, J. B. Fonseca-Neto, J. R. Nascimento, A. Yu.
Petrov, J. Ricardo}

\affiliation{Departamento de Física, Universidade Federal da Paraíba\\
 Caixa Postal 5008, 58051-970, Jo\~{a}o Pessoa, Para\'{i}ba, Brazil}

\email{pporfirio,jfonseca,jricardo,jroberto,petrov@fisica.ufpb.br}

\author{A. F. Santos}\email[]{alesandroferreira@fisica.ufmt.br}
\affiliation{Instituto de F\'{\i}sica, Universidade Federal de Mato Grosso,\\
78060-900, Cuiab\'{a}, Mato Grosso, Brazil}
\affiliation{Department of Physics and Astronomy, University of Victoria,\\
3800 Finnerty Road, Victoria, BC, Canada}

\begin{abstract}
We verify the consistency of the G\"{o}del-type solutions within the four-dimensional
Chern-Simons modified gravity with the non-dynamical Chern-Simons
coefficient, for different forms of matter including dust, fluid,
scalar field and electromagnetic field, and discuss the related causality issues.  We show that, unlike the general relativity, a vacuum solution is possible in our theory. Another essentially new result of our theory having no analogue in the general relativity consists in the existence of the hyperbolic causal solutions for a physically well-motivated matter.
\end{abstract}
\maketitle

\section{Introduction}

The studies of the Chern-Simons (CS) modified gravity have a long
story. Initially, the CS extension for the three-dimensional gravity has been proposed
in \cite{DJT} as an analogue for the CS extension of quantum electrodynamics
which allowed to construct the gauge invariant massive theory. The
four-dimensional extension of the gravitational CS term was firstly
introduced in \cite{AlvWitten} within the context
of gravitational anomalies, where it was shown to be characterized
by an arbitrary pseudo-scalar function $\phi(x)$, that is, the so-called
CS coefficient. The new wave of interest to the four-dimensional CS modified gravity,
that is, the gravitational model whose action represents itself as
a sum of the usual Einstein-Hilbert action and the gravitational CS
term, has been inspired by the seminal paper \cite{JaPi}, where the
four-dimensional gravitational CS term was shown to
involve Lorentz symmetry breaking for a special form of the CS coefficient
 $\phi(x)=k_{\mu}x^{\mu}$, and, evidently, 
the presence of the gravitational CS term breaks the CPT symmetry for any $\phi(x)$.
Therefore, the CS modified gravity represents itself as a simplest example
of a Lorentz-CPT breaking extension of the general relativity (GR).
Also, it was shown in \cite{JaPi} that the Schwarzschild metric
continues to be a solution for equations of motion also within CS modified gravity.
Further, the gravitational CS term has been discussed as a possible
ingredient of the most generic Lorentz-breaking extension of the standard
model \cite{Kostel}, and it was shown to arise as a quantum correction in the theory describing spinor fields on a curved background \cite{qugra}. Among other results found for the CS modified gravity in dimensions higher than three, one should also mention the possible higher-dimensional generalizations
of the gravitational CS term performed in \cite{Salgado}.  Different issues related to Chern-Simons (super)gravities in different dimensions has been discussed \cite{JZ}. A review on specifics of the CS gravity in four-dimensions has been developed \cite{Alexander}. 

It is clear that to study the compatibility of any new extended gravity
model, in particular, the CS modified gravity, with the observations, 
one should verify the consistency of the known solutions of the Einstein-Hilbert
gravity within the new theory. A wide class of the classical
gravitational solutions possessing spherical or axial symmetry has
been discussed from this viewpoint in the paper \cite{Grumiller}.
Besides of this, a further generalization of the CS modified gravity
has been proposed in that paper, that is, the so-called dynamical Chern-Simons
modified gravity where the CS coefficient $\phi(x)$ becomes a dynamical
field. This extension opened the possibilities for new studies. In
particular, it was showed in \cite{Konno} that the Kerr metric itself
does not solve the modified Einstein equations in this theory, and, to achieve consistency,
one should modify some elements of the metric tensor by additive terms depending on the CS coefficient. The cosmological solutions in the dynamical CS modified gravity
have been considered in \cite{Soda}.

It is
well known that the problem of causality in the gravitational context
is strongly polemical. Historically, this problem began to be discussed in \cite{Godel} where the G\"{o}del metric  allowing for the closed time-like curves (CTCs) was found.
Further, a very important generalization
of the G\"{o}del metric has been introduced, with a number of their properties
were studies, in a sequence of papers \cite{Reb}. In particular,
it was showed in these papers that in certain cases the CTCs are ruled
out for some special relations between constant parameters of this
metric. Therefore, it is interesting to verify the consistency of
the Gödel-type metric proposed in \cite{Reb} within the CS modified
gravity, especially, to obtain the conditions for parameters allowing
to rule out the CTCs. Some preliminary studies in this direction have
been performed in \cite{ourGodel}, where, first, only the case
of the original G\"{o}del metric has been considered, second, the results
seem to be dependent on the choice of the basis, which naturally calls the interest to solving this problem in terms of the tetrad basis.

Therefore, the natural problem would consist, first, in studying of
the compatibility of different kinds of the G\"{o}del-type metric (it
is known \cite{Reb} that there are three kinds of this metric, that
is, hyperbolic, trigonometric and linear cases) within the CS modified
gravity following the procedure used in \cite{Reb}, second, in determining values of the parameters which could rule out the CTC solutions. Namely
these problems are considered in this paper.

The structure of the paper looks like follows. In section II, we present a general review of the G\"{o}del-type metrics. Section III is devoted to description of the G\"{o}del-type solutions in the general relativity for different forms of the matter including a relativistic fluid, a cosmological term, scalar and electromagnetic fields, with use of the tetrad basis. In section IV, these solutions are generalized for the CS modified gravity, with the relations between parameters of matter are found explicitly. Finally, in Summary the results are discussed.

\section{G\"{o}del-type Metrics }

In this section, we briefly review the G\"{o}del-type metrics consistent with the condition of  homogeneity
in the space-time (ST-homogeneous), and discuss their classes and causality features
related to the existence of closed time-like curves. 

In spite of the local validity of the causality principle in general
relativity, assured by its local Lorentzian character, it is well
known that there are solutions of Einstein field equations displaying causality anomalies given by closed time-like curves. One
of the first and best known examples of such a solution is the famous G\"{o}del metric
\cite{Godel}. The rotating G\"{o}del universe belongs to a family
of solutions of Einstein equations which are ST-homogeneous, known
as G\"{o}del-type metrics \cite{Reb}, whose line element in cylindrical
coordinates is given by
\begin{equation}
ds^{2}=-[dt+H(r)d\theta]^{2}+D^{2}(r)d\theta^{2}+dr^{2}+dz^{2}\label{goty_ds2},
\end{equation}
and the metric functions $H(r)$ and $D(r)$ obey the following necessary
and sufficient conditions for ST-homogeneity
\begin{equation}
\begin{array}{ccc}
\frac{H'(r)}{D(r)} & = & 2\omega,\\
\frac{D''(r)}{D(r)} & = & m^{2},
\end{array}\label{eq:st_hom}
\end{equation}
where the prime denotes the derivative with respect to $r$. All the
G\"{o}del-type ST-homogeneous metrics are completely characterized by
the two parameters $(m^{2},\omega)$, where $\omega\neq0$ is the vorticity,
and $-\infty\leq m^{2}\leq\infty$, such that isometric space-times
have identical pairs ($m^{2},\omega^{2})$. The G\"{o}del-type ST-homogeneous
manifolds admit a $G_{5}$ group of isometries when $0\leq m^{2}<4\omega^{2}$,
a $G_{7}$ when $m^{2}=4\omega^{2}$ and a $G_{6}$ for the degenerate
case $\omega=0$ \cite{Reb}. There are three different classes
of non-degenerate ($\omega\neq0$) G\"{o}del-type metrics, according to
the solutions of Eqs. (\ref{eq:st_hom}), namely, 

i)\textit{ hyperbolic class}, where $m^{2}>0$ and
\begin{equation}
\begin{array}{cl}
H(r) & =\frac{2\omega}{m^{2}}[\cosh(mr)-1],\\
D(r) & =\frac{1}{m}\sinh(mr);
\end{array}\label{eq:goty_hip}
\end{equation}

ii)\textit{ trigonometric class}, where $m^{2}=-\mu^{2}<0$ and
\begin{equation}
\begin{array}{cl}
H(r) & =\frac{2\omega}{\mu^{2}}[1-\cos(\mu r)],\\
D(r) & =\frac{1}{\mu}\sin(\mu r);
\end{array}\label{eq:goty_trig}
\end{equation}

iii)\textit{ linear class}, where $m^{2}=0$ and
\begin{eqnarray}
H(r) & = & \omega r^2,\label{eq:goty_lin}\\
D(r) & = & r.\nonumber 
\end{eqnarray}
The G\"{o}del solution \cite{Godel} belongs to the hyperbolic class
in which $0\leq m^{2}=2\omega^{2}$. It is a solution of Einstein
equations with the cosmological constant $\Lambda$ for a dust of density
$\rho$ and rigid rotation $\omega$, such that $m^{2}=2\omega^{2}=-2\Lambda=\kappa\rho$,
where $\kappa$ is the Einstein constant. But through the change of
variables $\rho=p'-\rho'$, $\Lambda=\Lambda'-\kappa p'$ it can be
reinterpreted as a solution with a cosmological constant $\Lambda'$
and a perfect fluid with pressure $p'$, density $\rho'$ and vorticity
$\omega$, where $m^{2}=2\omega^{2}=\kappa(p'+\rho')$ and $2\Lambda'=p'-\rho'$.

The closed time-like curves of G\"{o}del-type space-times are circles
$C=\lbrace(t,r,\theta,z);\,t,r,z=const,\,\theta\in[0,2\pi]\rbrace$,
in a region defined by a range of $r$ ($r_{1}<r<r_{2}$) where $G(r)=D^{2}(r)-H^{2}(r)<0$.
For the linear class $m=0$, there is a non-causal region $r>r_{c}$
with closed time-like curves, where $r_{c}=1/\omega$ is a critical
radius. For the trigonometric class $m^{2}=-\mu^{2}<0$, there is
an infinite sequence of alternating causal and non-causal regions.
For the hyperbolic class $m^{2}>0$, when $0<m^{2}<4\omega^{2}$ there
is a non-causal region $r>r_{c}$, where the critical radius $r_{c}$
is given by
\begin{equation}
\sinh^{2}(\dfrac{mr_{c}}{2})=(\dfrac{4\omega^{2}}{m^{2}}-1)^{-1},\label{eq:raiocr}
\end{equation}
and when $m^{2}\geq4\omega^{2}$ there is no breakdown of causality,
since for $m^{2}=4\omega^{2}$ the critical radius $r_{c}\rightarrow\infty$
\cite{Reb}. There is no analogous situation for the linear and trigonometric
classes, where there is no way to circumvent the violation of causality.

\section{G\"{o}del-type Metrics in GR}

In this section we review the G\"{o}del-type solutions of GR for the physically
well-motivated matter content discussed in Refs. \cite{Reb}, having in mind our objective
to investigate their possible extensions for the case of
the CS modified gravity. In order to  characterize completely the G\"{o}del-type
universes in GR, it is necessary the consider the matter sources of
the G\"{o}del-type metrics. Since these metrics are ST-homogeneous,
there is a local Lorentz co-frame $w^{A}=e_{\;\;\mu}^{A}dx^{\mu}$
where the components of the curvature tensor are constants. It
is given by \cite{Reb}
\begin{equation}
\begin{split}w^{(0)} & =dt+H(r)d\theta,\\
w^{(1)} & =dr,\\
w^{(2)} & =D(r)d\theta,\\
w^{(3)} & =dz,
\end{split}
\label{eq:Lor_tetr}
\end{equation}
where $ds^{2}=\eta_{AB}w^{A}w^{B}$ and $\eta_{AB}=diag(-1,+1,+1,+1)$
is the Minkowski metric. The Einstein field equations
$R_{AB}-\frac{1}{2}\eta_{AB}R=\kappa T_{AB}-\eta_{AB}\Lambda$, in
the Lorentz frame (\ref{eq:Lor_tetr}), can be written as
\begin{equation}
\begin{split}R_{AB} & =\kappa(T_{AB}-\frac{1}{2}\eta_{AB}T)+\Lambda\eta_{AB},\end{split}
\label{eq:ein_eq}
\end{equation}
where $T=\eta^{AB}T_{AB}$ is the trace of energy-momentum tensor
of matter $T_{AB}$. 

Our conventions for the curvature and Ricci tensors are $R_{\;\mu\beta\nu}^{\alpha}=\Gamma_{\beta\mu,\nu}^{\alpha}-\Gamma_{\mu\nu,\beta}^{\alpha}+\Gamma_{\rho\nu}^{\alpha}\Gamma_{\beta\mu}^{\rho}-\Gamma_{\rho\beta}^{\alpha}\Gamma_{\nu\mu}^{\rho}$
and $R_{\mu\nu}=R_{\;\mu\alpha\nu}^{\alpha}.$ 
Small Greek letters
label coordinate indices and capital Latin letters label Lorentz tetrad
indices, and both take values from $0$ to $3$. 

For all three classes of G\"{o}del-type metrics given by Eqs. (\ref{eq:goty_hip}-\ref{eq:goty_lin}),
the curvature scalar is given by $R=2(\omega^{2}-m^{2})$, and the
non-vanishing components of Ricci tensor, in the Lorentz co-frame
defined by Eqs.(\ref{eq:Lor_tetr}) above, are $R_{(0)(0)}=2\omega^{2}$,
$R_{(1)(1)}=R_{(2)(2)}=2\omega^{2}-m^{2}$. Since $R_{AB}$ is constant
and diagonal, it follows that $T_{AB}$ has to be constant and diagonal,
due to the Einstein field equations. There is a preferred direction
in G\"{o}del-type space-times determined by the rotation vector which,
for the Local Lorentz frame observers defined by Eqs.(\ref{eq:Lor_tetr})
and 4-velocity $u^{A}=e_{\;\;0}^{A}=\delta_{\;\;0}^{A}$, is given
by $\omega^{A}=\frac{1}{2}\epsilon^{ABCD}u_{B}\omega_{CD}=[0,0,0,-\omega]$,
where $\epsilon^{ABCD}=\frac{1}{\sqrt{-g}}\varepsilon^{ABCD}$ ($\varepsilon^{0123}=1$)
is the totally antisymmetric Levi-Civita tensor, and $\omega_{AB}$
is the vorticity tensor. 

In order to compare the results of GR and CS modified gravity, we use the
same matter content of the universe considered in \cite{Reb}, that
is, a perfect fluid, a scalar field and an electromagnetic field.
Let us start with the perfect fluid, which has rigid rotation velocity
$\omega$, density $\rho$ and pressure $p$, whose components of
the energy-momentum tensor $T_{\quad AB}^{(pf)}=(p+\rho)u_{A}u_{B}+p\,\eta_{AB}$,
in the Lorentz tetrad basis defined by Eqs. (\ref{eq:Lor_tetr}), are
\begin{equation}
T_{\quad(0)(0)}^{(pf)}=\rho,\,\,T_{\quad(1)(1)}^{(pf)}=T_{\quad(2)(2)}^{(pf)}=T_{\quad(3)(3)}^{(pf)}=p,
\end{equation}
where the 4-velocity of an element of fluid is $u^{A}=e_{\;\;0}^{A}=\delta_{\;\;0}^{A}$.
Notice that $T_{\quad AB}^{\mathrm{(pf)}}$ is constant and diagonal.
The next matter source is a massless scalar field $\psi$ satisfying the Klein-Gordon equation $\square\psi=\eta^{AB}\nabla_{A}\nabla_{B}\psi=0$.
Following \cite{Reb}, we choose a scalar field whose gradient $\partial_{A}\psi=e_{A}^{\,\,\,\,\,\mu}\partial_{\mu}\psi$,
in the Lorentz tetrad basis defined by Eqs. (\ref{eq:Lor_tetr}), is
constant and parallel to $z$ axis. Thus, $\psi=s(z-z_{0})$, where $s$
is a constant. The components of its energy-momentum tensor $T_{\quad AB}^{(sf)}=\partial_{A}\psi\partial_{B}\psi-\frac{1}{2}\eta_{AB}\eta^{CD}\partial_{C}\psi\partial_{D}\psi$,
in the Lorentz co-frame given by Eqs. (\ref{eq:Lor_tetr}), are
\begin{equation}
T_{\quad(0)(0)}^{(sf)}=T_{\quad(3)(3)}^{(sf)}=\frac{s^{2}}{2},\,\,T_{\quad(1)(1)}^{(sf)}=T_{\quad(2)(2)}^{(sf)}=-\frac{s^{2}}{2}.
\end{equation}
Finally, the last source in \cite{Reb} is a source-free electromagnetic
field $F_{AB}$ that is a solution of the vacuum Maxwell equations
$\nabla\!_{B}F^{AB}=0$ and $\nabla_{\!B}{}^{\ast}\!F{}^{AB}=0$,
where $^{\ast}\!F{}^{AB}=\frac{1}{2}\epsilon^{ABCD}F_{CD}$. According
to \cite{Reb}, let us choose an electromagnetic field $F_{AB}$ such
that, in the Lorentz co-frame (\ref{eq:Lor_tetr}), both the electric
and the magnetic fields are parallel to $z$ axis, that is, $F_{(0)(3)}=E(z)=e\,\sin[2\omega(z-z_{0})]$
and $F_{(1)(2)}=B(z)=-e\,\cos[2\omega(z-z_{0})]$, where $e$ is a
constant. The non-vanishing components of the energy-momentum tensor
$T_{\quad AB}^{(ef)}=F_{A}^{\:\;N}F_{BN}-\frac{1}{4}\eta_{AB}F^{CD}F_{CD}$
of the electromagnetic field, in the Lorentz co-frame (\ref{eq:Lor_tetr}),
are given by
\begin{equation}
T_{\quad(0)(0)}^{(ef)}=T_{\quad(1)(1)}^{(ef)}=T_{\quad(2)(2)}^{(ef)}=\frac{e^{2}}{2},\,\,T_{\quad(3)(3)}^{(ef)}=-\frac{e^{2}}{2}.
\end{equation}
To resume, the energy-momentum tensor of all matter sources is
given by
\begin{equation}
T_{AB}=T_{\quad AB}^{(pf)}+T_{\quad AB}^{(sf)}+T_{\quad AB}^{(ef)},\label{eq:emom ten}
\end{equation}
which is diagonal with components constants. Thus, by taking the Einstein
constant $\kappa=1$, the Einstein equations (\ref{eq:ein_eq})
for an arbitrary G\"{o}del-type metric with parameters $(\omega,m^{2})$,
in the Lorentz tetrad basis (\ref{eq:Lor_tetr}), are
given by
\begin{eqnarray}
2\omega^{2} & = & \frac{1}{2}\,{\it e}^{2}+\frac{1}{2}\,\rho+\frac{3}{2}\,p-\Lambda,\label{eq:ein_eq00}\\
2\,\omega^{2}-m^{2} & = & \frac{1}{2}\,{\it e}^{2}-\frac{1}{2}\,p+\frac{1}{2}\,\rho+\Lambda,\label{eq:eineq11}\\
0 & = & -\frac{1}{2}\,{\it e}^{2}+s^{2}-\frac{1}{2}\,p+\frac{1}{2}\,\rho+\Lambda.\label{eq:ein_eq33}
\end{eqnarray}
The cosmological constant is completely determined by the matter source,
according to Eq. (\ref{eq:ein_eq33}), and does not depend on the metric
parameters $(\omega,m^{2})$, since $R_{(3)(3)}=0$. Solving the field
equations (\ref{eq:ein_eq00}--\ref{eq:ein_eq33}) for the cosmological
constant $\varLambda$ and the metric parameters $m^{2}$ and $\omega$,
we obtain that
\begin{eqnarray}
2\omega^{2} & = & \rho+p+s^{2},\label{eq:gr_2w2}\\
m^{2} & = & \rho+p+2s^{2}-e^{2},\label{eq:gr_m2}\\
2\Lambda & = & -\rho+p-2s^{2}+e^{2}.\label{eq:gr_2L}
\end{eqnarray}

This matter content gives rise to solutions belonging to all three
classes of G\"{o}del-type metrics, but both the linear and trigonometric
classes require the existence of the electromagnetic field, according
to Eq. (\ref{eq:gr_m2}). Solutions for the hyperbolic class, where
$m^{2}>0$, are obtained when $\rho+p+2s^{2}-e^{2}>0$. There is an
upper bound for $m^{2}$ given by $m^{2}\leq4\omega^{2}$, since
\begin{equation}
m^{2}-4\omega^{2}=-(p+\rho+e^{2})\leq0.\label{eq:gr_causal}
\end{equation}
Therefore, there is no solution within the causal region, except the
Rebou\c{c}as-Tiomno solution, where $m^{2}=4\omega^{2}$, for a matter described purely by the scalar
field \cite{Reb}. The G\"{o}del solution $m^{2}=2\,\omega^{2}=3\,s^{2}+p+\rho-2\,e^{2}>0$
is recovered either for all the sources, when $s^{2}=e^{2}\neq0$,
or for a pure perfect fluid, when $s^{2}=e^{2}=0$, since
\begin{equation}
m^{2}-2\omega^{2}=s^{2}-e^{2}.\label{eq:gr_godel}
\end{equation}
Furthermore, when $s^{2}\leq e^{2}$, there is a lower upper bound
for $m^{2}$ given by $m^{2}\leq2\omega^{2}$, and when $s^{2}\geq e^{2}$,
there is a non-zero lower bound given by $2\omega^{2}\leq m^{2}$. 

The solutions for the linear class, where $m^{2}=0$, are obtained
when $\rho+p+2s^{2}-e^{2}=0$, and the solutions of the trigonometric
class, where $m^{2}=-\mu^{2}<0,$ are obtained when $\rho+p+2s^{2}-e^{2}<0$.
An important result is that the perfect fluid and the scalar field produce
hyperbolic solutions within the interval $2\omega^{2}\leq m^{2}\leq4\omega^{2}$. 

The sign of the cosmological constant for all three classes can be
easily determined from
\begin{equation}
2\Lambda=2p-m^{2},\label{eq:gr_Lsign}
\end{equation}
which is obtained by summing of Eqs. (\ref{eq:gr_2L}) and (\ref{eq:gr_m2}).
For the hyperbolic class, the cosmological constant is positive when
$2p>m^{2}>0$, negative when $0\leq2p<m^{2}$ and zero when $0<2p=m^{2}$.
For the linear class, we have $\Lambda=p\geq0$, since $m^{2}=0$,
and for the trigonometric class, where $m^{2}=-\mu^{2}<0,$ we have
only $2\Lambda=2p+\mu^{2}>0$. 

Note that the presence of the electromagnetic field is essential for
the consistency of all three classes of solutions, in other words, in its absence, one has only the hyperbolic class, where
all solutions are within the interval $0<m^{2}<4\omega^{2}$ inside
the non-causal region, except of a pure scalar field corresponding to the case $m^{2}=4\omega^{2}$.

\section{G\"{o}del-type Metrics in Chern-Simons Modified Gravity}

In this section,  after a brief review of the non-dynamical CS
modified gravity, we investigate the existence of G\"{o}del-type
ST-homogeneous solutions of the CS-modified gravitational field equations,
as well as their causality features, in order to compare the GR and
CS gravitational theories with respect to this kind of causality violations.
In particular, since we are searching for generalizations of the G\"{o}del-type
solutions obtained in \cite{Reb}, the same matter content
will be the source for the CS modified gravitational field.

The CS modified gravity action \cite{JaPi} with the cosmological
constant is 
\begin{equation}
S=\frac{1}{2\kappa}\int d^{4}x\left[\sqrt{-g}(R-2\Lambda)+\frac{1}{4}\phi\:^{\ast}\!RR\right]+S_{m},\label{eq:cs_action}
\end{equation}
where $\phi$ is a scalar field, $S_{m}$ is the matter action and
$^{\ast}\!RR$ is the Pontryagin density defined by 
\begin{equation}
\begin{split}^{\ast}\!RR & \equiv{}^{\ast}\!R_{\;\;B}^{A\;\;CD}R_{\ ACD}^{B}=\frac{1}{2}\varepsilon^{CDEF}R_{\ \ BEF}^{A}R_{\ ACD}^{B}\end{split}
.\label{eq:potryangin}
\end{equation}
The scalar field $\phi$ is called the CS coefficient, or CS coupling
field, and measures how the modified theory deforms the GR.

The CS gravitational field equations are obtained by varying the CS
modified action with respect to the metric $g_{\mu\nu}$ and the CS
coupling field $\phi$ and, in the Lorentz tetrad basis
defined by Eqs. (\ref{eq:Lor_tetr}), they can be written as
\begin{eqnarray}
R_{AB}+C_{AB} & = & \kappa(T_{AB}-\frac{1}{2}\eta_{AB}T)+\eta_{AB}\Lambda,\label{eq:cseq_met}\\
^{*}\!RR & = & 0,\label{eq:cseq_phi}
\end{eqnarray}
where $T$ is the trace of the matter energy-momentum tensor $T^{AB}$,
and $C^{AB}$ is the Cotton tensor arising due to variation of the CS term,
defined as \cite{JaPi}: 
\begin{eqnarray}
C^{AB} & = & -\frac{1}{2}\bigg[(\epsilon^{CADE}\nabla_{D}R_{\;\:E}^{B}+\epsilon^{CBDE}\nabla_{D}R_{\;\:E}^{A})\partial_{C}\phi\label{eq:c_ten}\\
 &  & +(^{\ast}\!R^{EAFB}+^{\ast}\!R^{EBFA})\nabla_{E}\nabla_{\!F}\phi\bigg]\,.\nonumber 
\end{eqnarray}
The Eq. (\ref{eq:cseq_phi}) means that the theory involves a constraint,
i.e., such a field equation imposes restrictions on the possible space-time
geometries. Furthermore, the covariant divergence of the Cotton tensor
is 
\begin{equation}
\nabla_{A}C^{AB}=\frac{1}{8\sqrt{-g}}\:{}^{\ast}\!RR\,\partial^{B}\phi,\label{eq:div c_ten}
\end{equation}
that implies, by Eqs. (\ref{eq:cseq_met}), which the covariant divergence
of $T_{AB}$ is proportional to $^{*}\!RR$, in other words, the Pontryagin
density measures the rate of breaking of the diffeomorphism invariance.

We shall study the G\"{o}del-type metrics in the CS modified gravity
context and investigate the causality features, as well as the possible
simultaneous solutions among GR and the CS modified gravity. For this,
we shall consider the same matter sources used in the paper \cite{Reb}
and combinations between them as particular cases.

In order to calculate the CS gravitational field equations for a G\"{o}del-type
metric, it is necessary to calculate first the Cotton tensor given by Eqs. (\ref{eq:c_ten})
and the Pontryagin scalar density given by Eq. (\ref{eq:potryangin}). For an
arbitrary CS coupling field $\phi(t,r,z)$ with a cylindrical symmetry
and an arbitrary G\"{o}del-type metric with components $H(r)$ and $D(r)$,
we obtain that $^{*}\!RR=0$, and that the non-vanishing components
of the Cotton tensor are given by the diagonal components
\begin{equation}
C_{(0)(0)}=2C_{(1)(1)}=2C_{(2)(2)}=2\frac{\partial\phi}{\partial z}\,\omega(4\omega^{2}-m^{2}),\label{eq:cten_diag}
\end{equation}
and the non-diagonal components
\begin{eqnarray}
C_{(0)(1)} & = & -\frac{1}{2}\frac{\partial^{2}\phi}{\partial z\partial t}\,\frac{H(r)}{D(r)}\,(4\omega^{2}-m^{2}),\nonumber \\
C_{(0)(2)} & = & -\frac{1}{2}\frac{\partial^{2}\phi}{\partial z\partial r}\,(4\omega^{2}-m^{2}),\nonumber \\
C_{(0)(3)} & = & -\frac{\partial\phi}{\partial t}\,\omega(4\omega^{2}-m^{2}),\label{eq:cten_ndiag}\\
C_{(1)(3)} & = & -\frac{1}{2}\frac{\partial^{2}\phi}{\partial t^{2}}\,\frac{H(r)}{D(r)}\,(4\omega^{2}-m^{2}),\nonumber \\
C_{(2)(3)} & = & \frac{1}{2}\frac{\partial^{2}\phi}{\partial t\partial r}\,(4\omega^{2}-m^{2})\,.\nonumber 
\end{eqnarray}

The Rebou\c{c}as-Tiomno G\"{o}del-type universe, corresponding to $4\omega^{2}=m^{2}$,
solves the CS gravitational field equations trivially, since it
follows from Eqs. (\ref{eq:cten_diag}-\ref{eq:cten_ndiag}) that in this case
$C_{AB}=0$, for an arbitrary CS coupling field $\phi$. Therefore,
for $4\omega^{2}=m^{2}$ the CS field equations are identical to
Einstein field equations, and we must require that $4\omega^{2}\neq m^{2}$
in order to obtain nontrivial solutions of the CS modified gravitational field equations. 

Taking into account that for an arbitrary ST-homogeneous G\"{o}del-type metric
the Ricci tensor is diagonal, and that the matter source considered
in \cite{Reb} is also diagonal in the Lorentz tetrad basis
defined by Eqs. (\ref{eq:Lor_tetr}), one concludes that the non-diagonal
elements of the Cotton tensor (\ref{eq:cten_ndiag}) should vanish.
Therefore, we have the equations $C_{AB}=0$ (for $A\neq B$), whose solution
is $\phi(t,r,z)=b(z-z_{0})+f(r)$, where $b$ and $z_{0}$ are arbitrary
constants and $f(r)$ is an arbitrary function. Since the diagonal
components of the Cotton tensor (\ref{eq:cten_diag}) involve only
the derivatives $\frac{\partial\phi}{\partial z}$, we can choose
$f(r)=0$ for the sake of simplicity without loss of generality. Therefore the CS coupling field
is restricted to be
\begin{equation}
\phi(z)=b(z-z_{0}),\label{eq:phisol}
\end{equation}
whose gradient $\partial_{A}\phi=[0,0,0,b]$ is parallel to the vorticity
vector $\omega_{A}=[0,0,0,-\omega]$ and aligned with the $z$ axis,
the preferred direction of the ST-homogeneous G\"{o}del-type metrics.
This allows us to introduce a new arbitrary parameter $k$ defined
by the inner product $\omega^{A}\partial_{A}\phi=-b\,\omega=-k$.
Both the gradient $\partial_{A}\phi$ of the CS scalar field and the vorticity
$\omega^{A}$ have the same direction and sense when $k<0$.
Thus, since there is a coupling between the vorticity and the CS scalar
field, we can choose either $b$ or $k$ as an independent parameter.
As we will see later, it is better to consider 
\begin{equation}
b=\frac{k}{\omega}.\label{eq:cs_bsol}
\end{equation}

In order to compare the results of GR and CS gravity, we continue to use the matter
content of the G\"{o}del-type solutions of GR obtained in \cite{Reb},
whose energy-momentum tensor is given by Eqs. (\ref{eq:emom ten}),
as well the CS modified gravitational field
equations (\ref{eq:cseq_met}). Thus, for the CS scalar field given
by Eq. (\ref{eq:phisol}) and an arbitrary G\"{o}del-type metric with components
$H(r)$ and $D(r)$, in the Lorentz tetrad basis defined by Eqs. (\ref{eq:Lor_tetr}),
we obtain the following form of the CS modified gravitational field equations
\begin{eqnarray}
2\,\omega^{2}+2\,b\omega(4\,\omega^{2}-m^{2}) & = & \frac{1}{2}\,e^{2}+\frac{1}{2}\,\rho-\Lambda+\frac{3}{2}\,p,\label{eq:cs00}\\
2\,\omega^{2}-m^{2}+b\omega(4\,\omega^{2}-m^{2}) & = & \frac{1}{2}\,e^{2}-\frac{1}{2}\,p+\Lambda+\frac{1}{2}\,\rho,\label{eq:cs22}\\
0 & = & -\frac{1}{2}\,e^{2}-\frac{1}{2}\,p+s^{2}+\Lambda+\frac{1}{2}\,\rho,\label{eq:cs33}
\end{eqnarray}
with the requirement that $4\omega^{2}\neq m^{2}$, in order to have
$C_{AB}\neq0$ for an arbitrary CS coefficient.

In CS modified gravity the cosmological constant $\Lambda$ is related only
to the matter content and is the same for both GR and CS solutions,
given by Eq. (\ref{eq:gr_2L}), that is, $2\Lambda=-\rho+p-2s^{2}+e^{2}$,
since the CS field equation (\ref{eq:cs33}) is equal to the
equation (\ref{eq:ein_eq33}). The remaining two CS gravitational equations
(\ref{eq:cs00}-\ref{eq:cs22}) can be used to determine the
metric parameters $(\omega,m^{2})$ in terms of the matter content
and the CS scalar field, where $b$ is considered as an arbitrary
parameter, since there are three parameters and two equations. Following
this procedure, expressing $m^2$ from (\ref{eq:cs22}) and substituting
in Eq. (\ref{eq:cs00}) we arrive at a third order algebraic equation
for $\omega$. On the other hand, taking into account that the CS scalar field
parameter $b$ appears in the CS gravitational equations only through the
combination $b\omega$, we substitute Eq. (\ref{eq:cs_bsol}) in the equations for
CS gravitational field (\ref{eq:cs00}-\ref{eq:cs22}). 
Treating $k$ as an arbitrary parameter, we obtain linear equations
with respect to both $m^{2}$ and $\omega^{2}$ which, after substituting
$\Lambda$ through (\ref{eq:gr_2L}), are given by
\begin{eqnarray}
\left(2+8\,k\right)\omega^{2}-2\,km^{2} & = & \rho+s^{2}+p,\label{eq:csk00}\\
\left(2+4\,k\right)\omega^{2}-\left(1+k\right)m^{2} & = & -s^{2}+e^{2}.\label{eq:csk22}
\end{eqnarray}

Although there is no Gödel-type vacuum solution in GR, the CS modified field
equations (\ref{eq:csk00}-\ref{eq:csk22}) have a vacuum solution
given by
\begin{eqnarray}
\Lambda & = & 0,\nonumber \\
m^{2} & = & \omega^{2},\label{eq:cs_vac}\\
b & = & \frac{k}{\omega}=-\frac{1}{3\,\omega}\nonumber 
\end{eqnarray}
This solution belongs to the hyperbolic class and
presents causal anomalies, since $0<m^{2}<4\omega^{2}$, therefore we see that the causality violation can occur without any matter, only due to the CS coefficient itself which can be treated as a some special non-dynamical media. Even being
non-dynamical, the CS field $\phi$ interacts with the metric in such
a way that the vacuum CS field equations give rise to a non-trivial
space-time geometry, where the non-vanishing components of the Ricci
tensor are $R_{(0)(0)}=2\,R_{(1)(1)}=2\,R_{(2)(2)}=2\,\omega$ and
the curvature scalar is $R=0$. 

On the other hand, the non-vacuum
solution of the CS field equations (\ref{eq:csk00}-\ref{eq:csk22})
for $\omega^{2}(k)$ is given by 
\begin{equation}
\omega^{2}(k)=\frac{a_{\omega}k+b_{\omega}}{3k+1}=\frac{1}{2}\frac{(\rho+p+3\,s^{2}-2\,e^{2})\,k+\rho+p+s^{2}}{3k+1},\label{eq:cs_w2}
\end{equation}
for $k\neq-\frac{1}{3}$, whose coefficients
\begin{eqnarray}
2\,a_{\omega} & = & \rho+p+3\,s^{2}-2\,e^{2},\label{eq:aw2}\\
2\,b_{\omega} & = & \rho+p+s^{2},\label{eq:bw2}
\end{eqnarray}
are determined only by the matter content. Similarly the solution for $m^{2}(k)$
is given by 
\begin{equation}
m^{2}(k)=\frac{a_{m}k+b_{m}}{3k+1}=\frac{2\,(\rho+p+3\,s^{2}-2\,e^{2})\,k+\rho+p+2\,s^{2}-e^{2}}{3k+1},\label{eq:cs_m2}
\end{equation}
for $k\neq-\frac{1}{3}$, with coefficients 
\begin{eqnarray}
a_{m} & = & 2\,(\rho+p+3\,s^{2}-2\,e^{2}),\label{eq:am2}\\
b_{m} & = & \rho+p+2\,s^{2}-e^{2}.\label{eq:cs_bm2}
\end{eqnarray}

The signs of the metric parameters $\omega^{2}$ and $m^{2}$
given by the GR solution (\ref{eq:gr_2w2}-\ref{eq:gr_2L}), depend on the matter content and on the range of $k$.
Since $\omega$ is real, we must impose the condition $\omega^{2}(k)>0$,
which determines the allowed range of $k$ and, in turn, also determine
the possible signs of $m^{2}(k)$ and the classes of the Gödel-type
metric within the allowed range. 

When the CS coefficient vanishes, i.e. $k=0$, or, as is the same, $b=0$, see (\ref{eq:cs_bsol}),
the CS solution given by (\ref{eq:cs_w2}-\ref{eq:cs_m2}) reduces to $\omega^{2}=b_{\omega}$
and $m^{2}=b_{m}$, thus reproducing the GR solution (\ref{eq:gr_2w2}-\ref{eq:gr_2L}).
In other words, the coefficients $b_{\omega}$ and $b_{m}$ of the
CS solution are reduced to the GR results for $\omega^{2}$ and $m^{2}$, respectively,
for the same matter content. 

The causality features of the CS solution  (\ref{eq:cs_w2}-\ref{eq:cs_m2})
is determined by the sign of
\begin{equation}
m^{2}-4\,\omega^{2}=\frac{-(\rho+p+e^{2})}{3k+1},\label{eq:cs_causal}
\end{equation}
which depends on the allowed range of $k$. In general, for $k<-\frac{1}{3}$
we have $m^{2}>4\,\omega^{2}$ and the solution is causal, while for $k>-\frac{1}{3}$
we have $m^{2}<4\,\omega^{2}$ and the solution is non-causal. And
for $k=-\frac{1}{3}$, we have the vacuum solution which is non-causal.

There are general relations between the coefficients of $\omega^{2}$
and $m^{2}$ above, given by $a_{m}=4\,a_{\omega}$ and
$b_{m}=b_{\omega}+a_{\omega}$. The root $k_{\omega}=-\frac{b_{\omega}}{a_{\omega}}$
of $\omega^{2}(k)$ and the root $k_{m}=-\frac{b_{m}}{a_{m}}$ of
$m^{2}(k)$ are also related as $k_{\omega}=4\,k_{m}+1$. There are
also the restrictions ${\it a_{\omega}}-3\,{\it b_{\omega}}=-\rho-p-e^{2}\leq0$
and ${\it a_{m}}-3\,{\it b_{m}}=-\rho-p-e^{2}\leq0$ . Using the conditions
$a_{\omega}=3\,b_{\omega}$ and $a_{m}=3\,b_{m}$ in the CS solution,
which takes place only for a pure scalar field source $\psi=s\,(z-z_{0})$,
we obtain that $\omega^{2}(k)=b_{\omega}=\frac{1}{2}s^{2}$ and $m^{2}(k)=b_{m}=2\,s^{2}$
(for any $k$) . This is the Rebou\c{c}as-Tiomno GR solution \cite{Reb}, where $m^{2}=4\,\omega^{2}$, for which the Cotton tensor vanishes, independently of the CS scalar field. Thus,  differently from GR, CS gravity
has only a trivial solution in presence of a pure scalar field. 

\begin{figure}[h]
\includegraphics[scale=0.4]{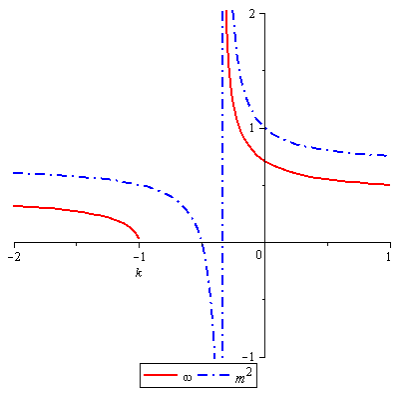}

\caption{\label{fig:cs_wm2_dust}CS solution $\omega(k)$ and $m^{2}(k)$ for
dust with $\rho=1$. }
\end{figure}

There is a special CS solution obtained when the matter source
satisfies the conditions $2\,a_{\omega}=\rho+p+3\,s^{2}-2\,e^{2}>0$ and
$2\,b_{\omega}=\rho+p+s^{2}>0$, since these conditions imply that
both $a_{m}>0$ and $b_{m}>0$ and the roots of $\omega^{2}(k)$
and $m^{2}(k)$ are related through the relation $k_{\omega}=-\frac{b_{\omega}}{a_{\omega}}<k_{m}=-\frac{b_{m}}{a_{m}}<-\frac{1}{3}$.
The main features of this solution is most evident when the source
is given by a pure dust, so that CS solution (\ref{eq:cs_w2}-\ref{eq:cs_m2})
can be written as 
\begin{eqnarray}
\frac{\omega^{2}(k)}{\rho} & = & \frac{1}{2}\,\frac{k+1}{3\,k+1},\label{eq:cs_w2_dust}\\
\frac{m^{2}(k)}{\rho} & = & \frac{2\,k+1}{3\,k+1},\label{eq:cs_m2_dust}
\end{eqnarray}
whose roots are $k_{\omega}=-1$ and $k_{m}=-1/2$, respectively.
This dust solution must satisfy the requirement $\omega^{2}(k)>0$,
since $\omega$ is real, which determines the allowed range of $k$
given by $k<-1$ or $k>-1/3$, where $m^{2}(k)>0$ as well. This
can be easily seen in Fig. (\ref{fig:cs_wm2_dust}), with the graphics
of the $\omega(k)>0$ and $m^{2}(k)$, are given for $\rho=1$. Therefore, the
G\"{o}del-type metric of the CS dust solution above is hyperbolic and
has a GR analogue, since $k=0$ belongs to the
allowed range. Treating the causality, for $k<-1$ the space-time is causal and
for $k>-1/3$ the space-time is non-causal. A distinguishing feature
of CS gravity, contrasting to GR, is the existence of the hyperbolic causal solutions
for physically well-motivated matter sources, whose energy-momentum
tensor given by Eq. (\ref{eq:emom ten}), where $\rho+p+3\,s^{2}-2\,e^{2}>0$.

The G\"{o}del metric solution of CS gravity can be obtained by imposing,
for the CS solution (\ref{eq:cs_w2}-\ref{eq:cs_m2}) the following condition
\begin{equation}
m^{2}-2\,\omega^{2}=-\frac{\left(-3\,s^{2}-p-\rho+2\,e^{2}\right)k+e^{2}-s^{2}}{3\,k+1}=0,
\end{equation}
whose result can be presented as
\begin{eqnarray}
m^{2} & = & 2\,\omega^{2}=3\,s^{2}+p+\rho-2\,e^{2}>0,\label{eq:cs_wm_god}\\
b\omega & = & k=\frac{e^{2}-s^{2}}{2\,\omega^{2}}\neq0.\label{eq:cs_bk_god}
\end{eqnarray}
Therefore, it is insufficient to have a pure perfect fluid to achieve this solution, as happens
in GR, since the condition $b\neq0$ requires the presence either of an electromagnetic
field or of a scalar field, according to Eq. (\ref{eq:cs_bk_god}). On
the other hand, the matter source can also be given by an electromagnetic field
and a scalar field, where $e^{2}-s^{2}\neq0$. This is an extension
of the GR solution, which is recovered when $e^{2}-s^{2}=0$.

Now let us discuss the existence of the linear and trigonometric
classes in CS modified gravity. Without the electromagnetic field, all solutions
have the same behaviour as the dust solution (\ref{eq:cs_w2_dust}-\ref{eq:cs_m2_dust}),
where all coefficients $a_{\omega}$, $b_{\omega}$ , $a_{m}$ and
$b_{m}$ are positive, belonging only to the hyperbolic class. Therefore,
similarly to GR, the presence of the electromagnetic field is necessary in order to
obtain solutions of linear and trigonometric classes. Although the GR
solution given by Eqs. (\ref{eq:gr_2w2}-\ref{eq:gr_2L}),
for the non-degenerate case where $\omega\neq0$, cannot be achieved if the source is presented by a
pure electromagnetic field, for the CS solution given by
Eqs. (\ref{eq:cs_w2}-\ref{eq:cs_m2}) and Eq. (\ref{eq:gr_2L}),
this is not only possible, but also all the classes are obtained.

\begin{figure}[h]
\includegraphics[scale=0.5]{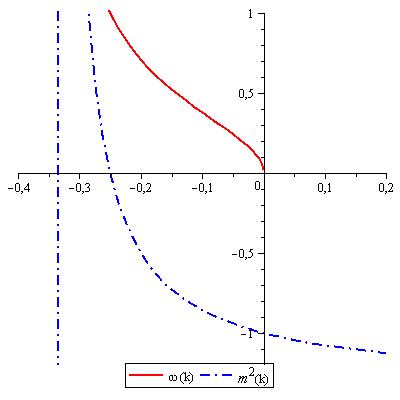}\caption{\label{fig:wm2_e}$w$ and $m^2$, for $e=1$}
\end{figure}

The CS solution given by Eq. (\ref{eq:gr_2L}) and Eqs. (\ref{eq:cs_w2}-\ref{eq:cs_m2}),
when the source is a pure electromagnetic field, is given by $2\Lambda=e^{2}$
and
\begin{eqnarray}
\frac{\omega^{2}}{e^{2}} & =- & \frac{k}{3k+1},\label{eq:cs_w2_e}\\
\frac{m^{2}}{e^{2}} & =- & \frac{4\,k+1}{3k+1},\label{eq:cs_m2_e}
\end{eqnarray}
respectively. The roots of $\omega^{2}$ and $m^{2}$ are $k_{\omega}=0$
and $k_{m}=-1/4$, correspondingly, and do not depend on the value of
$e^{2}$. This solution must satisfy the
requirement $\omega^{2}(k)>0$, since $\omega$ is real. It follows
that the allowed range of $k$ is given by $-1/3<k<0$ and that the
solution given by the Eqs. (\ref{eq:cs_w2_e}-\ref{eq:cs_m2_e})
is hyperbolic ($m^{2}>0$) for $-1/3<k<-1/4$, linear ($m^{2}=0$)
for $k_{m}=-1/4$ and trigonometric ($m^{2}<0$) for $-1/4<k<0$.
These results can be seen in Fig. \ref{fig:wm2_e}, with the graphics
of $\omega(k)$ and $m^{2}(k)$ are given for $e^{2}=1$, which shows the main
features of the pure electromagnetic CS solution. The hyperbolic solution
is non-causal, since $\omega^{2}>m^{2}>0$ for $-1/3<k<0$, that is,
in all allowed range of $k$. All of these solutions have no analogue
in GR.

\section{Summary}

We discussed the consistency of the G\"{o}del-type metric and the possibility of existence of closed time-like curves within the CS modified gravity. Within our studies, we
used a tetrad basis and a tetrad formalism allowing to simplify the calculations crucially, and considered the metric corresponding to a non-zero Cotton tensor, differently from \cite{ourGodel} where the consistency of the G\"{o}del metric has been verified only in the case of the vanishing Cotton tensor. We considered all three characteristic forms of a generic G\"{o}del-type metric, including not
only hyperbolic case but also trigonometric and linear cases, and  a rather generic form of the
matter, involving a relativistic fluid, a cosmological constant, scalar and electromagnetic fields.

We found that the G\"{o}del-type metric indeed solves the modified
Einstein equations in certain cases. The most interesting among them
is certainly the hyperbolic case which is reproduced if the CS
term is switched off. We also found that in certain cases we can have solutions
consistent with existence of CTCs. We found that unlike the GR, in the CS modified gravity there is a nontrivial solution for the vacuum case. Also, we found several other solutions having no analogues in the GR case, with the most important among them is the hyperbolic causal solution.

We demonstrated that the GR limit in the CS modified gravity is recovered by making the amplitude of CS coefficient
equal to zero, $b=0$. Furthermore, due to arbitrariness of parameter $b$, it opened up a wide range of the G\"{o}del-type possible solutions, among which we obtained some hyperbolic causal solutions whose existence is impossible
in GR, except of pure scalar field case. We found that the presence of the electromagnetic field plays the crucial role for linear and trigonometric solutions, just as in the usual GR case. Thus, we conclude that the
impact of the CS extension becomes fundamental for the causality features for all relevant matter sources. 
We note that arising of all these solutions does not require any presence of the exotic matter.

\textbf{Acknowledgments.} Authors are grateful to M. Rebou\c{c}as for important discussions. This work was partially supported by Conselho
Nacional de Desenvolvimento Científico e Tecnológico (CNPq). The work
by A. Yu. P. has been supported by the CNPq project No. 303783/2015-0. The work by A. F. S. is supported by CNPq projects 476166/2013-6 and 201273/2015-2.

\end{document}